\documentstyle[editedvolume,numreferences]{crckapb}

\begin{opening}
\title{Effects of atomic interactions 
in two component submonolayer growth}

\author{Miroslav Kotrla}
\institute{Institute of Physics
           Academy of Sciences of the Czech Republic\\
          Na Slovance 2, 182 21 Praha 8, Czech Republic}
\author{Joachim Krug}
\institute{Fachbereich Physik, Universit\"at Essen \\ D-45117 
Essen, Germany}

\end{opening}

\runningtitle{TWO COMPONENT SURFACE GROWTH}

\begin{document}


\begin{abstract}
We discuss effects of the different chemical bondings 
on the island morphology and on the island density scaling 
in two-component submonolayer growth. 
Different regimes depending on  the strength of the mutual 
interactions and on the relative mobility of species 
are described
and studied by kinetic Monte Carlo
simulations using a two-component solid-on-solid growth model.
Results for the temperature and flux dependence
of the island density as well as examples of the surface morphologies
are presented.
\end{abstract}

\section{Introduction}
Recent progress in both experimental techniques and theoretical modeling
of epitaxial growth allowed to improve our understanding of growth
mechanisms on the atomic scale.
In particular, a satisfactory understanding of
growth in homoepitaxial systems is beginning
to emerge.
Many results were accumulated by basic material-oriented research
as well as by general studies in the spirit of statistical
mechanics.
For example, the processes of nucleation and aggregation in single component
submonolayer growth \cite{brune98} 
are now well understood.

On the other hand, practically important materials prepared
by growth contain several components or impurities.
Our present understanding
of the real microscopic processes going on 
in such systems is rather incomplete.
The investigation of multicomponent growth is more complicated
not only because we have to consider
more material parameters than in homoepitaxy, 
but because there are additional
possible atomistic processes which have to be considered
(inlayer or interlayer exchange, floating 
of adsorbates on the surface or the island edges, additional energy barriers
due to island or step edge decoration  etc.).

Multicomponent growth is controlled by the interactions between species.
The interactions are determined 
by the chemical nature of components, as well as
by the geometrical positions of the atoms. 
There are two basic features of submonolayer multicomponent growth which affect
the surface morphology: effects of the strain due to atomic mismatch,
and effects of the different chemical bondings between components of 
the system. 
Although it is broadly accepted that the presence of strain plays a 
crucial role in heteroepitaxy, in particular in the formation of
ordered nanostructures, the
different chemical bondings by themselves 
can result in interesting effects. 
As an example we may mention Hwang's
STM study of growth of 
Co on Ru(0001) with predeposited Ag \cite{hwang96}.
He observed the formation of dendritic islands, 
despite the fact that neither of 
the two component metals exhibit this behavior. 
Hwang suggested that this can be explained 
by chemically induced step edge diffusion barriers between domains of 
different species,
which should generally exist in multicomponent film growth.
Another example are the effects of phase separation
in multilayer molecular beam epitaxy \cite{leo_lar_des_97}.

In this paper, we concentrate on effects
of the difference in chemical bondings between the atoms of 
a two-component system growing in the
submonolayer regime. 
We discuss the effects on the resulting island morphology,
and address the question of
the modification of island density scaling.

We present some general considerations on different growth
regimes in two component growth in Section \ref{sec:regimes}.
In the next Section \ref{sec:model}, 
we describe a model for
submonolayer two-component growth.
Section  \ref{sec:scaling} very briefly recalls the scaling law
for the island density.
In section  \ref{sec:adsorbates}, we discuss the case of a small
concentration of one component.
Section  \ref{sec:alloy} 
contains results of kinetic Monte Carlo simulations
of two component alloy growth.
In particular, we show how the scaling of the 
island density is modified by the 
relative mobility
of the components and interactions between them.

\section{Different regimes of multicomponent growth}
\label{sec:regimes}
Growth in the presence of more than one type of particles is realized
in various ways.
The simplest scenario is the
deposition of a certain type of
particles on a substrate composed from a different material.
This is usually called  heteroepitaxy.
Interesting effects like growth-induced surface alloying
of species which are immiscible in the bulk have been observed
in such systems.

There are many experimental studies of
growth in which two or more different components were deposited
on a substrate consisting in general of a different material 
(see e.g. \cite{ven97}).
In the case of multicomponent growth, we need to distinguish
different regimes of growth. The species can be either {\it codeposited}
simultaneously with generally different fluxes, 
or deposited sequentially, one of the
species being {\it predeposited}.
We restrict ourselves in this paper to the simplest situation 
of two component growth.
We call the components {\it A} and {\it B}, and we consider
the situation where an alloy consisting of {\it A} and {\it B}
, or a material ({\it A})
in the presence of impurities ({\it B}) grows on a certain substrate.

Important material parameters in submonolayer
growth are the strengths of the interactions,
both laterally and perpendicular to the surface.
In comparison to homoepitaxy, two-component growth
contains many more material parameters. However, 
within a certain temperature 
window only some of them may be important.
The crucial parameter is
the interaction between particles of different type.
It can be both attractive or repulsive. We consider here
the former.
Depending on the relative strengh of this interaction with respect
to the interactions between particles of the same type, one can 
observe {\it intermixing} (possibly leading to antiferromagnetic-like ordering)
or {\it phase separation}.
The second important parameter is the relative mobility of the different
species. 

To simplify the classification, we fix one lateral interaction,
let us say between {\it A} and {\it A}, and the
mobility of one species, let us say of {\it A}.
The interaction between {\it A} and {\it B} and 
the interaction between {\it B} and {\it B}        
can be stronger or weaker
than the interaction between {\it A} and {\it A}, and 
the mobility of {\it B} can be higher or lower than the mobility of {\it A}.
This yields 12 qualitatively different situations of submonolayer
growth in a bicomponent system.
(In the case of the multilayer growth, such classification
cannot be easily done due to the presence of 
additional interlayer processes.)
It is important to note that the precise values of parameters
needed to decide in which regime growth proceeds are often
not well known. In this situation, it seems to be useful to explore
different possible effects on a general level. 
We are not going to analyze all the situation here but
we discuss some partial results.

\section{Two-component submonolayer growth model}
\label{sec:model}
To carry out an effective analysis we need a reasonably simple, yet 
flexible model. Kinetic Monte Carlo (KMC) simulations proved to 
be a useful tool for evaluating the importance of various microscopic 
processes. 
We use the variant of the solid-on-solid model
with two surface species described in detail in \cite{kks00b}. The
basic microscopic processes are random deposition and migration.
Two types of particles are deposited with fluxes $F_A$ and $F_B$.
In the case of codeposition the concentration
of the two components is changed by 
the variation of the ratio $F_A/F_B$.

Diffusion is modeled by hopping to nearest-neighbor sites
with the rate
$R_D^X=k_0 \exp (-E_D^X /k_B T)$
for a particle of the type $X = A$ or $B$.
The energy barrier $E_D^X=  E_{\rm sub}^{X}
+ \sum_{Y=A,B} n^{XY} E_{\rm n} ^{XY}$  
depends on the interaction with the substrate (the first term)
and on the lateral neighbors (the second term).

The interaction with the substrate is
affected by the particle beneath the particle under consideration.
It may be a substrate particle, which
can be in general of a type different from {\it A} or {\it B}
 (let us call it {\it C}),
or a previously deposited particle of type {\it A} or {\it B}.
We have in general six possibilities:
$E_{\rm sub}^{AC}$, $E_{\rm sub}^{BC}$,
$E_{\rm sub}^{AA}$, $E_{\rm sub}^{BA}$ $E_{\rm sub}^{AB}$, $E_{\rm sub}^{BB}$.
However, in the situation of a low coverage and fast descent to
the lower terrace one can suppose that 
only $E_{\rm sub}^{A} =  E_{\rm sub}^{AC}$ and
$E_{\rm sub}^{B} = E_{\rm sub}^{BC}$ are the relevant
two parameters controlling the mobility of both particle types.

The interaction with the nearest-neighbors is controlled by
bond counting, i.e.
$n^{XY}$ is the number of nearest-neighbor
$X$-$Y$ pairs and $E_{\rm n}^{XY}$ 
the corresponding contribution to the 
barrier, with $E_{\rm n}^{AB} = E_{\rm n}^{BA}$.
In all examples presented in this paper $E_{\rm n}^{AA} = 0.3$~eV, 
$E_{\rm sub}^{A} = 0.8$~eV and $k_0 = 10^{13}$ Hz.

\section{Island density scaling}
\label{sec:scaling}
We call the island density $N$.
In homoepitaxy, the scaling relation
\begin{equation}
N \sim (F/D)^\chi, ~~~~~~~~
\label{subMLscal}
\chi = \frac{i^\ast}{i^\ast+2}
\end{equation}
between the deposition flux $F$ and the adatom diffusion coefficient $D$
has been well established 
theoretically [6--8]
numerically  [8--12]
and experimentally 
\cite{stroscio94}. 
In (\ref{subMLscal}), 
$i^\ast$ denotes the size of the largest unstable cluster.  

Little is known about the 
details of nucleation during submonolayer growth
of two component systems.
The presence of the second species affects the nucleation of
the first one and vice versa. 
An obvious complication is that there are
different types of islands. In principle, 
one can consider three island
densities: $N_A$ ($N_B$) for islands composed only from {\it A} ({\it B}), and
$N$ for islands composed from both particle types.
In the case of mixing, there is no problem, since $N$ is the relevant quantity.
The situation is less clear
in the presence of phase
separation. Here islands composed only from one type of
particles may exist for weak or repulsive {\it A}-{\it B} interactions.  

\section{Effects of adsorbates}
\label{sec:adsorbates}

The situation is simpler if the coverage of one of the components
is much smaller than the total coverage 
(let us say $\theta_B \ll  \theta$, then the {\it B}-component is 
an impurity), or if there is another mechanism
like floating of one component ({\it B})
on the edges of the islands of the other component ({\it A}), since 
then $N_A \approx N$.

Several microscopic mechanisms by which impurities
could alter the relationship
(\ref{subMLscal}) were considered in the literature.
First, impurities may act as nucleation centers, thus
effectively {\em decreasing} $i^\ast$ and therefore $\chi$; in the
extreme case of immobile adatom traps, the limit of spontaneous
nucleation with $i^\ast = \chi = 0$ would be 
realized \cite{amar95,chambliss94}. 
Recently, we have shown that the predeposition of
only a low concentration of impurities
($\theta_B$ a few thousandth of ML) can lead to a severalfold increase of the
island density \cite{kks00c}.
However, the increase significantly depends on the
mobility of the adsorbates. In the case of essentially 
immobile adsorbates ($E_{\rm sub}^{B}=5$ eV),
a new feature in the flux dependence of the island
density appears. Instead of a single power law relationship,
there is a plateau where $N \approx \theta_B$,
reflecting the dominance of heterogeneous nucleation  
in a certain flux \cite{kks00c} (temperature \cite{kks01a}) interval.
When its origin is not recognized, the evaluation of such a plateau
within the usual analysis of island density scaling \cite{brune98}
may lead to an anomalously low estimate of the adatom diffusion barrier,
as well as to an anomalously small preexponential factor. This effect
has been invoked \cite{michely01} to explain the extremely low diffusion
prefactors reported recently for several metal surfaces \cite{barth00}.   

Second, impurities decorating the island
edges may induce energy barriers to attachment. Kandel \cite{kandel97}
predicted using a rate equation theory that, 
provided these barriers are sufficiently strong,
the exponent $\chi$ in (\ref{subMLscal}) is {\em increased\/} such that
the above expression is replaced by $\chi = 2 i^\ast/(i^\ast + 3)$.
Also this mechanisms implies an 
increase of the island density compared to the case of pure
homoepitaxy.
We found in KMC simulation 
that the key kinetic process for growth of decorated islands
is the activated exchange of impurities and adatoms \cite{kks00a}.
The rate of this process is taken as
$k_{\rm ex}=k_0 \exp (- E_{\rm ex}/k_B T)$,
where $E_{\rm ex}$ is the corresponding activation barrier.
In this simulation the barriers for exchange 
$E_{\rm ex}$ and for the {\it B}-mobility $E_{\rm sub}^{B}$ 
are fixed at $E_{\rm ex} = E_{\rm sub}^{B} = 1$~eV.
We observed well decorated islands for both 
$E_{\rm n}^{AB} < E_{\rm n}^{AA}$ and
$E_{\rm n}^{AB} > E_{\rm n}^{AA}$ (left panel of Fig. 1).
We found that, in our case \cite{kks00b}, the floating mobile adsorbates
indeed strongly increase the island density but
without appreciably changing its power-law dependence on the flux
(right panel of Fig. 1).
The increase is stronger for larger $E_{\rm n}^{AB}$, and
only slightly higher for predeposition than
for codeposition. 
We also observed a stronger coverage dependence of the island density
in comparison with homoepitaxy (right panel of Fig. 1).
This was interpreted as 
a delay of the saturation regime, where the
island density becomes independent of coverage. 
The coverage dependence is again more 
pronounced for strongly interacting adsorbates,
but it is weaker in the case of predeposition than for codeposition.

\begin{figure}[ht]
\centering
\vspace*{64mm}
\label{fig1}
\includegraphics{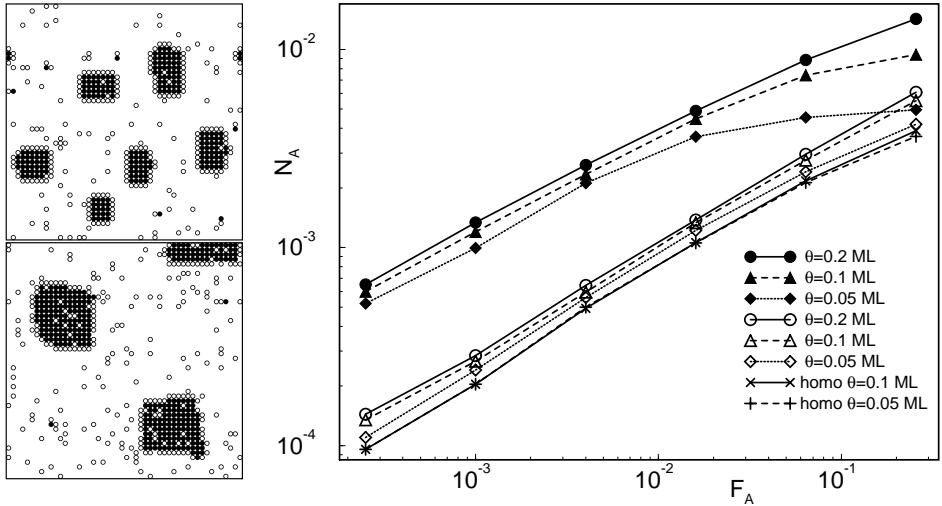}
\caption{Left panels: Examples of configurations 
obtained by codeposition with fluxes 
$F_A = F_B = 0.004$~ML/s at temperature $T=500$~K
 using parameters:
$E_{\rm ex}= 1$~eV,
and  $E_{\rm n}^{AB} = 0.4$~eV (upper subpanel),  
$E_{\rm n}^{AB} = 0.2$~eV (lower subpanel).
Closed (open) circles represent {\it A} ({\it B})
atoms. 
Coverage $\theta =  \theta_A + \theta_B = 0.2$~ML.
We show only $50 \times 50$ sections of larger simulation boxes.
Right panel: Averaged island density $N_A$ as a function of flux
$F_A$ in codeposition with $F_B = F_A$ 
 for several values of the total coverage 
$\theta$
and different  energy barriers:
$E_{\rm n}^{AB} = 0.2$~eV - open symbols,   
$E_{\rm n}^{AB} = 0.4$~eV - filled symbols.
The exchange barrier $E_{\rm ex}=1$~eV.
The behavior in homoepitaxy ($F_B = 0$) is shown for comparison.} 
\end{figure}
 
Third, it has been argued on the basis of KMC simulations \cite{liu95}
that \emph{repulsive} impurities can alter the exponent $\chi$ in 
(\ref{subMLscal}) by rendering the diffusion of the adatoms anomalous,
in the sense that their mean square displacement behaves as 
$\langle x^2 \rangle \sim t^{\alpha}$ with $\alpha < 1$. A scaling 
analysis for $i^\ast = 1$ then predicts that $\chi = \alpha/(1 + 2 \alpha)
< 1/3$, i.e. the island density scaling exponent is predicted to 
decrease in this case.

\section{Island density in binary alloy}
\label{sec:alloy}
In this section, we study nucleation 
in a situation where the concentration of both species is comparable.
We shall discuss effects of the different mobility of both components
and of the variation of the {\it A}-{\it B} and {\it B}-{\it B}
 interactions; as before,
the mobility of {\it A}-atoms and {\it A}-{\it A} interaction is fixed:
$E_{\rm sub}^{A} = 0.8$ eV,   $E_{\rm n}^{AA} = 0.3$ eV .
We consider (i) the regime 
of phase separation,
  $E_{\rm n}^{AB} = 0.2$~eV 
and   $E_{\rm n}^{BB} = 0.4$ eV, and 
(ii) the regime of the intermixing, 
  $E_{\rm n}^{AB} = 0.4$ eV  and $E_{\rm n}^{BB} = 0.2$ eV.
In all simulations the particles are codeposited
and the results are analyzed at the total coverage $\theta=0.1$~ML.

\begin{figure}[ht]
\centering
\vspace*{56mm}
\label{fig2}
\includegraphics{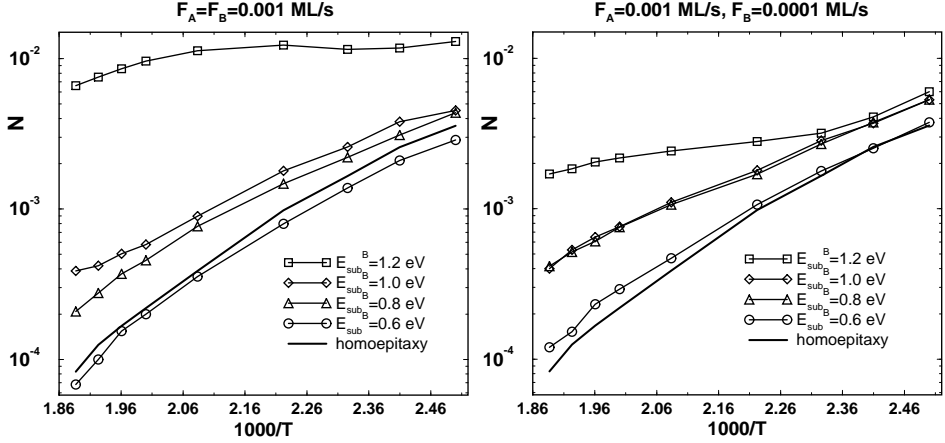}
\caption{Averaged density $N$ of islands composed from both types of particles
as a function of the inverse temperature
for different mobilities of {\it B}-component
in the regime of phase separation for codeposition
with fluxes
$F_A = F_B = 0.001$~ML/s (left panel) and
$F_A = 0.001$~ML/s,  $F_B = 0.0001$~ML/s (right panel), $\theta = 0.1$~ML.
The behaviour in homoepitaxy, the case when there is no {\it B}-component,  
is shown for comparison.}
\end{figure}

In Fig. 2,
we show the island density $N$ as a function of the inverse temperature
for different mobilities of the {\it B}-component
in the regime of phase separation for two concentrations
$F_A = F_B = 0.001$~ML/s (left panel) and
$F_A = 0.001$~ML/s,  $F_B = 0.0001$~ML/s (right panel).
The behaviour in homoepitaxy, the case when there is no {\it B}-component,  
is shown for comparison.
One can see that there is a substantial increase of the island
density. Similar as for the case of immobile impurities,
there is the indication of a plateau
for the case of the lowest mobility of the {\it B}-component
for both concentrations.

Hence, the mobility of the atoms of one component strongly influences
the island density of the growing alloy even its concentration is small.
Note that if the mobility of {\it B} is higher than the mobility of {\it A},
then the island density is \emph{lower} than in homoepitaxy
for $F_A = F_B = 0.001$~ML/s,
but not for $F_A = 0.001$~ML/s,  $F_B = 0.0001$~ML/s.
If one analyzes separately the density $N_A$ of islands composed only from
{\it A} (not shown) no clear picture appears.

The change of the island density with mobility is accompanied
by a change of the island morphology. This is illustrated in 
the left column of Fig. 3
showing configurations at $T=500$~K
for different mobilities of {\it B}-atoms
($F_A = F_B = 0.001$~ML/s).
We can see a transition from a situation in which 
{\it A}-islands are incorporated
with comparably large {\it B}-islands into 
\newpage
\begin{figure}[th]
\centering
\label{12conf}
\vspace*{159mm}
\includegraphics{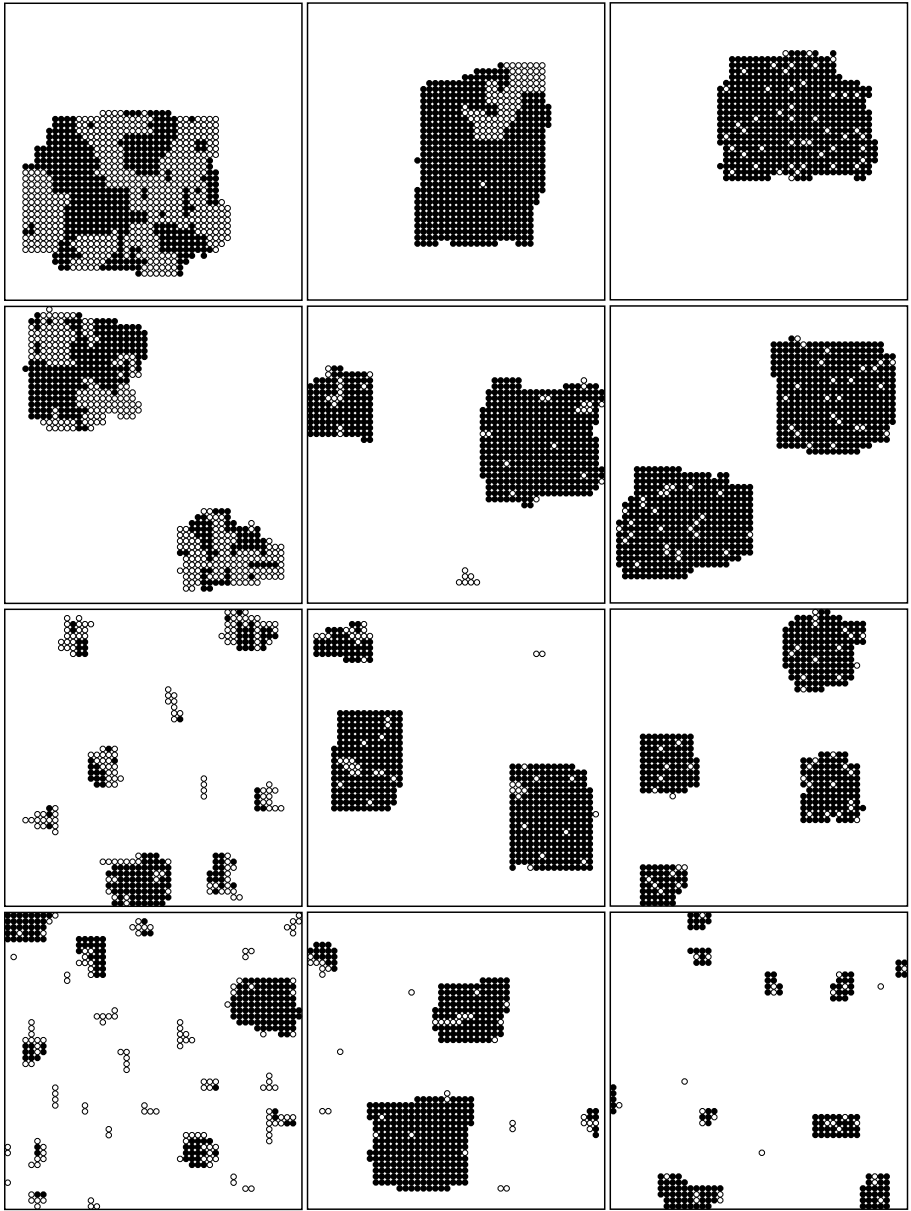}
\caption{
Examples of configurations for different {\it B}-mobility 
from top to bottom
  $E_{\rm sub}^{B} = 0.6$~eV,
  $E_{\rm sub}^{B} = 0.8$~eV,
  $E_{\rm sub}^{B} = 1.0$~eV,
  $E_{\rm sub}^{B} = 1.2$~eV
and various regimes:
phase separation
with fluxes $F_A = F_B =0.001$~ML/s 
(left column)
and
with fluxes $F_A = 0.001$~ML/s, $F_B =0.0001$~ML/s 
(middle column),
and intermixing
with fluxes $F_A = 0.001$~ML/s, $F_B =0.0001$~ML/s 
 (right column).
Closed (open) circles represent {\it A} ({\it B})
atoms. 
The total  coverage $\theta = 0.1$~ML, temperature $T=500$ K.
We show only 50 $\times$ 50 sections of larger simulation
boxes.
}
\end{figure}

\begin{figure}[ht]
\centering
\vspace*{56mm}
\label{fig4}
\includegraphics{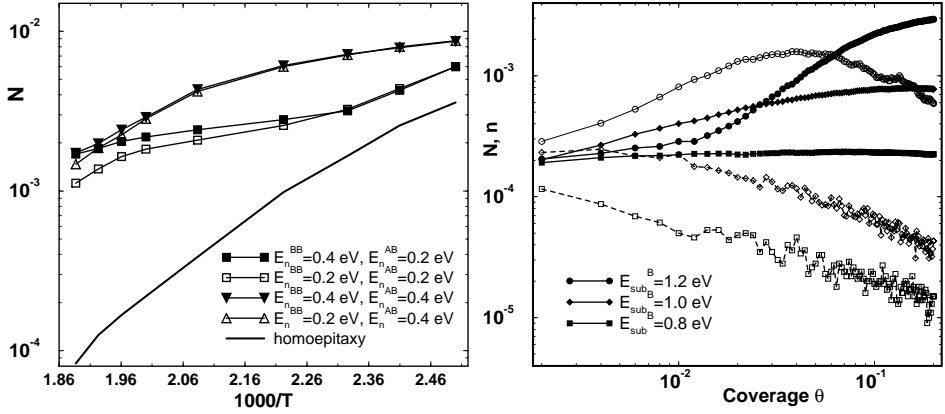}
\caption{Temperature (left panel) and coverage dependences
(right panel) for codeposition with different fluxes $F_A=0.001$~ML/s 
and $F_B = 0.0001$~ML/s. 
Left panel: density of islands $N$
as a function of the inverse temperature
for low mobility of {\it B} component,  $E_{\rm sub}^{B} = 1.2$ eV 
and different energy barriers $E_{n}^{BB}$ and  $E_{n}^{AB}$,
coverage is $\theta = 0.1$~ML.
Right panel: density of islands $N$ (filled symbols)
and free adatoms $n$ (open symbols)
as a function of coverage $\theta$
for three different energy barriers
  $E_{\rm sub}^{B} = 1.2$ eV (circles),
  $E_{\rm sub}^{B} = 1.0$ eV (diamonds)
and   $E_{\rm sub}^{B} = 0.8$ eV (squares),
$E_{n}^{BB}=0.4$~eV and  $E_{n}^{AB}=0.2$~eV, $T = 500$~K.} 
\end{figure}
\noindent
one island, to a 
situation in which there are {\it A}-islands attached to small {\it B}-islands.
The second column corresponds to the right panel in 
Fig. 2. 
The effect of {\it B}-mobility on the island morphology
(as well as on the island dendity cf. Fig. 2)
is now weaker.
The right column shows the situation in the intermixing regime.

The left panel of Fig. 4
compares island densities in
the regimes of phase separation and intermixing
for low {\it B} mobility and two {\it B}-{\it B} interaction strengths.
We can see that in the intermixing regime the presence of a
small fraction of a less mobile component increases the island density
more than in the phase separation regime (cf. the middle and the right
columns in Fig. 3). 
The increase of the strength of {\it B}-{\it B} interaction
has only a small effect, except for high temperatures in the phase
separation regime.

In the right panel of Fig. 4
we display
the coverage dependence of the island density and the density of
free adatoms in the phase separation regime for different mobilities of the
{\it B}-component.
There is a delay of the saturation in the island 
density and of decrease of the 
free adatoms density. The effect is stronger for lower {\it B}-mobility.
It can be explained by the continuous creation of centers for heterogeneous
nucleation.

\paragraph{Acknowledgements.}
We thank P. \v{S}milauer for useful discussion.
Support by the COST project P3.130 and
by Volkswagenstiftung is gratefully acknowledged.


\end{document}